\documentclass[twocolumn,preprintnumbers,amsmath,amssymb,showpacs]{revtex4}
 \usepackage{dcolumn}
 \usepackage{bm}

\begin{document}

 \preprint{}

 \title{Topological Properties of Phase Singularities in Wave Fields}

 \author {Yi-Shi Duan}
 \author{Ji-Rong Ren }
 \author {Tao Zhu }
 \thanks{Corresponding author. Email : zhut05@lzu.cn }
\affiliation{Institute of Theoretical Physics, Lanzhou University,
Lanzhou 730000, P. R. China}

 \date{\today}

 \begin{abstract}
Phase singularities as topological objects of wave fields appear in
a variety of physical, chemical, and biological scenarios. In this
paper, by making use of the $\phi$-mapping topological current
theory, we study the topological properties of the phase
singularities in two and three dimensional space in details. The
topological inner structure of the phase singularities are obtained,
and the topological charge of the phase singularities are expressed
by the topological numbers: Hopf indices and Brouwer degrees.
Furthermore, the topological invariant of the closed and knotted
phase singularities in three dimensional space are also discussed in
details.
\end{abstract}

 \pacs{02. 40. Xx,  03. 65. Vf,  05. 45. Yv}


 \maketitle

 \section{Introduction}

In physics, the solutions of wave equations in two and three
dimensional space often possess phase
singularities\cite{phasesingu,phase}. At the point of phase
singularity, the phase of the wave is undefined and wave intensity
vanishes. This phase singularity is a common property to all waves,
it exists in many area of physical, chemical, and biological
scenarios, such as the quantized vortices in superfluid or
superconductor systems\cite{supflu,quanvortex}, the vortices phase
singularities in Bose-Einstein condensates\cite{BEC}, the
streamlines or singularities in quantum mechanical
wavefunctions\cite{quanwave,quanvor}, the optical vortices in
optical wave systems\cite{opt1,opt2,opt3,optic,opttopo1,opttopo2},
the vortex filaments in chemical reaction and molecular diffusion
\cite{chemwave,chemwave2,chem}, and the phase singularities in
biological systems\cite{chem,bio}. In recent years, the phase
singularities have drawn great interest because it is of importance
for understanding fundamental wave physics and have many important
applications, and a great deal of works on the phase singularities
in wave fields have been done by many
physicists\cite{phasesingu,phase,opttopo1,opttopo2}.

Phase singularities are generically points in two dimensional space
and form a network of lines in three dimensional space. Around the
phase singularity points in two dimensions or phase singularity
lines in three dimensions, the phase changes by $2\pi$ times an
integer. This integer is the so-called topological charge (the
strength of the singularity) of the phase singularity objects
(points or lines). In three dimensional space, a important case is
that the phase singularity lines are closed and knotted curves,
these knot-like configurations exist in a variety of physical
scenarios, including Bose-Einstein condensations\cite{becknot},
chemical reaction and molecular diffusion
systems\cite{chemwave,chemwave2,chem}, optical wave
systems\cite{optknot1,optknot2} and field theory\cite{feild,field2}.
The phase singularities (points, lines, or knotted) are topological
objects, and have a very rich topological properties, so it is
necessary to use the topological method to study the topology of
these objects.

In this paper, we have given a generic feature of the topological
properties of the phase singularities in wave fields. By making use
of the $\phi$-mapping topological current theory\cite{topoduan}, we
study the topological current of the phase singularity objects
(points in two dimensions and lines in three dimensions), and the
topological inner structure of these topological objects are
obtained. For the case that the phase singularity lines in three
dimensional space are closed and knotted  curves, we discussed the
topological invariant of these knotted family in details. This paper
is arranged as follows. In Sec.II, we construct a topological
current density of the phase singularity points in two dimensions,
this topological current density don't vanish only when the phase
singularity points exist, the topological charge of phase
singularity points are expressed by the topological quantum
numbers,the Hopf indices and Brouwer degrees of the $\phi$-mapping.
In Sec.III is the topology of the phase singularity lines in three
dimensions. In Sec.IV, we introduce a important topological
invariant to describe the phase singularity lines when they are
linked and knotted, it is just the sum of all the self-linking and
all the linking numbers of the knot family. In Sec.V is our
concluding remarks.

\section{Topological structure of phase singularities in two dimensions}

Let us study a complex scalar field $\psi(\vec{x},t)$, the wave
fields function $\psi(\vec{x},t)$ is maps from space (of either two
or three dimensions) to the complex numbers, so $\psi : R^2,
R^3\rightarrow \mathcal {C}$. The complex scalar field
$\psi(\vec{x},t)$ is time dependent as well as space dependent, and
it can be written as
\begin{equation}
\psi(\vec{x},t)=\|\psi\|~
e^{i\theta}=\phi^1(\vec{x},t)+i\phi^2(\vec{x},t),\label{wavefunction}
\end{equation}
where $\phi^1(\vec{x},t)$ and $\phi^2(\vec{x},t)$ can be regarded as
complex representation of a two-dimensional vector field
$\vec{\phi}=(\phi^1,\phi^2)$ over two or three dimensional space,
$\|\psi\|=\sqrt{\psi^*\psi}$ is the modulus of $\psi$, and $\theta$
is the phase factor. In this section, we will restrict our attention
to the phase singularities in two dimensional space, and labeled
points $\vec{x}=(x^1, x^2)$ in cartesian coordinates. The phase
singularities in three dimensional space will be discussed in
Sec.III and Sec.IV.

It is known that the current density $\vec{J}$ associated with
$\psi(\vec{x},t)$ is defined as
\begin{equation}
\vec{J}=Im (\psi^* \nabla \psi)=\|\psi\|^2 \vec{v},\label{current}
\end{equation}
the $\vec{v}$ is the velocity field. From the expressions in
Eq.(\ref{wavefunction}) and Eq.(\ref{current}), the velocity field
$\vec{v}$ can be rewritten as
\begin{equation}
\vec{v}=\frac{1}{2i}\frac{1}{\|\psi\|^2}(\psi^*\nabla\psi-\nabla\psi^*\psi)=\nabla\theta,\label{velocity}
\end{equation}
it becomes the gradient of the phase factor $\theta$. The vorticity
field $\vec{\omega}$ of the velocity field $\vec{v}$ is defined as
\begin{equation}
\vec{\omega}=\frac{1}{2\pi}\nabla\times\vec{v}.\label{vorticity}
\end{equation}
Obviously, in two dimensional space the $\vec{\omega}$ only has
$z$-component, i.e., $\vec{\omega}=\omega \cdot \vec{e}_z$. Form
Eq.(\ref{velocity}),we directly obtain a trivial curl-free result:
$\vec{\omega}=\frac{1}{2\pi}\nabla\times\nabla\theta=0$. But in
topology, because of the existence of  phase singularities in the
wave fields $\psi$, the vorticity $\vec{\omega}$ does not
vanish\cite{quanvortex}. So in the following discussions, we will
study that what the exact expression for $\vec{\omega}$ is in
topology.

Introducing the unit vector $n^a=\phi^a/\|\phi\| (a=1, 2;
n^an^a=1)$, and defining a topological current in (2+1) dimensions
space-time,
\begin{equation}
j^i=\frac{1}{2\pi}\epsilon^{ijk}\epsilon_{ab}\partial_jn^a\partial_kn^b,~~~~~~~~~~i=0,
1, 2.
\end{equation}
Obviously, the topological current is identically conserved,
\begin{equation}
\partial_ij^i=0.\label{conserved}
\end{equation}

By making use of the definition of the unit vector $n^a$, one can
reexpressed the velocity field $\vec{v}$ as
\begin{equation}
\vec{v}=\epsilon_{ab}n^a\nabla n^b,
\end{equation}
and the vorticity $\vec{\omega}$ is
\begin{equation}
\omega=j^0=\frac{1}{2\pi}\epsilon^{jk}\epsilon_{ab}\partial_jn^a\partial_kn^b~~~~~~j,k=1,2,\label{topovorticity}
\end{equation}
it is just the time component of the topological current $j^i$.
According to the $\phi$-mapping topological current
theory\cite{topoduan}, one can prove that
\begin{equation}
j^i=\delta^2(\vec{\phi})D^i(\frac{\phi}{x}),\label{TopoExpressCurrent}
\end{equation}
where
\begin{equation}
D^i(\frac{\phi}{x})=\frac{1}{2}\varepsilon^{ijk}\epsilon_{ab}\partial_j\phi^a\partial_k\phi^b
\label{Jacobian}
\end{equation}
is the Jacobian vector, in which the time component $D^0(\phi/x)$ is
the usual two dimensional Jacobian $D(\phi/x)$. The expression of
the topological current in Eq.(\ref{TopoExpressCurrent}) shows that
the topological current $j^i$ does not vanish only at the zero
points of $\vec{\phi}$, i.e., the phase singularities of the wave
fields $\psi$. According to the Eq.(\ref{topovorticity}) and
Eq.(\ref{TopoExpressCurrent}), the vorticity can be  expressed
\begin{equation}
\omega=j^0=\delta^2(\vec{\phi})D(\frac{\phi}{x}),\label{topovorticitya}
\end{equation}
obviously, $\omega$ also does not vanish only when the phase
singularities of the wave fields $\psi$ exist. The topological
current $j^i$ and vorticity $\omega$ paly an important role in the
topological property of the phase singularities, so it is necessary
to study the zero points of $\vec{\phi}$ to determine the non-zero
solutions of $j^i$.

According to the implicit function theory\cite{impfun}, while the
regular condition $$D^i(\phi/x)\neq0$$ is satisfied, the general
solutions of
\begin{equation}
\phi^1(x^1,x^2,t)=0,~~~~\phi^2(x^1,x^2,t)=0,\label{Zeros of
function}
\end{equation}
can be expressed as
\begin{equation}
x^1=x^1_k(s,t), ~~~x^2=x^2_k(s,t),  ~~~k=1, 2, \ldots,
N,\label{solutions}
\end{equation}
which represent $N$ zero points $\vec{z}_k$ (k=1, 2, \ldots, N)
where $\psi(\vec{r},t)=0$ in two-dimensional space. These zero
points solutions are just the so-called phase singularity points of
the wave fields $\psi$, they represent the zeroes of the real part
and imaginary part of the complex scalar wave fields:
$Re(\psi)=\phi^1=0$ and $Im(\psi)=\phi^2=0$.

In the theory of $\delta$ function of $\vec{\phi}(\vec{r})$, one can
prove that in 2-dimensional space\cite{deltfun}
\begin{equation}
\delta^2(\vec{\phi})=\sum_{k=1}^{N}\frac{\beta_k}{|D(\phi/x)_{\vec{z}_k}|}\delta^2(\vec{x}-\vec{z}_k),\label{delta}
\end{equation}
where the positive integer $\beta_k$ is called the Hopf index of map
$x\rightarrow\phi$\cite{opttopo1,opttopo2}. The meaning of $\beta_k$
is that when the point $\vec{z}_k$ covers the neighborhood
$\Sigma_k$ of zero $\vec{z}_k$ once, the vector field $\vec{\phi}$
covers the corresponding region $\beta_k$ times. By substituting
Eq.(\ref{delta}) into Eq.(\ref{TopoExpressCurrent}), one can obtain
that
\begin{equation}
j^i=\sum_{k=1}^NW_k \delta^2(\vec{x}-\vec{z}_k)\frac{dx^i}{dt},
\end{equation}
where $W_k=\beta_k\eta_k$ is the winding number of $\vec{\phi}$
around zero point $\vec{z}_k$, and $\eta_k=sgn(D(\phi/x))=\pm1$ is
the Brouwer degrees of $\phi$-mapping. The sign of Brouwer degrees
are very important, for the case of vortices phase singularities,
the $\eta_k=+1$ corresponds to the vortex, and $\eta_k=-1$
corresponds to the antivortex.

The topological charge (the strength of phase singularities) of the
phase singularities point $\vec{z}_k$ is defined by the Gauss map
$\vec{n} :
\partial\Sigma_k\rightarrow S^1$\cite{gaussmap}:
\begin{equation}
W_k=\frac{1}{2\pi}\int_{\partial\Sigma_k}n^*(\epsilon_{ab}n^adn^b)
\end{equation}
where $n^*$ is the pullback of Gauss map $n$, $\partial\Sigma_k$ is
the boundary of a neighborhood $\Sigma_k$ of point $\vec{z}_k$ and
$\Sigma_k\cap\Sigma_m=\varnothing$ for $\Sigma_m$ is the
neighborhood  of another arbitrary zero point $\vec{z}_m$. In
topology it means that, when the point $\vec{z}_k$ covers
$\partial\Sigma_k$ once, the unit vector $\vec{n}$ will cover $S^1$,
or $\vec{\phi}$ covers the corresponding region $W_k$ times, which
is a topological invariant. By using the Stokes' theorem and in term
of Eq.(\ref{topovorticity}), one can obtain that
\begin{eqnarray}
W_k&=&\frac{1}{2\pi}\int_{\partial\Sigma_k}\epsilon_{ab}
\epsilon^{jk}\partial_jn^a\partial_kn^bd^2x\nonumber\\&&=\int_{\partial\Sigma_k}\omega
d^2x=\beta_k\eta_k.
\end{eqnarray}
It is clear that the topological charges densities of phase
singularity points is just the non-vanishing vorticity $\omega$ of
the velocity field $\vec{v}$. Using the notion $j=(\omega,
\vec{j})$, Eq.(\ref{conserved}) can be rewrite as
\begin{equation}
\frac{\partial\omega}{\partial t}+\nabla\cdot\vec{j}=0,
\end{equation}
this expression tells us that the topological charges of the phase
singularity points are conserved. This is the only topological
property of the complex wave fields.

In this section, we obtain the topological inner structure of the
phase singularity points in two dimensional space, the solutions of
the Eq.(\ref{solutions}) possess $N$ isolated phase singularity
points of which the $k$th point possesses topological charge
$W_k=\beta_k\eta_k$.

\section{Topological current of phase singularities in three dimensions}
In this section,we will study the phase singularities in three
dimensions space, some results which discussed in the above section
in two dimensions space are also useful here. Differently with in
two dimensions, the point is labeled $\vec{x}=(x^1,x^2,x^3)$ and the
vorticity $\vec{\omega}$ have its all three components. In three
dimensional space, the $\vec{\omega}$ can be reexpressed as
\begin{equation}
\omega^i=\frac{1}{2\pi}\epsilon^{ijk}\epsilon_{ab}\partial_jn^a\partial_kn^b,
\end{equation}
it is clearly that the vorticity $\vec{\omega}$ is just the
topological current of phase singularities in three dimensions
space.

Using the $\phi$-mapping theory\cite{topoduan}, the topological
current $\vec{\omega}$ is rewritten as
\begin{equation}
\omega^i=\delta^2(\vec{\phi})D^i(\frac{\phi}{x}),
\end{equation}
we can see from this expression that the vorticity $\vec{\omega}$ is
non-vanishing only if $\vec{\phi}=0$, i.e., the existence of the
phase singularities, so it is necessary to study these zero
solutions of $\vec{\phi}$. In three dimensions space, these
solutions are some isolate zero lines, which are the so-called phase
singularity lines in three dimensions space.

Under the regular condition $$D^i(\phi/x)\neq0,$$ the general
solutions of
\begin{equation}
\phi^1(x^1,x^2,x^3,t)=0,~~~~~ \phi^2(x^1,x^2,x^3,t)=0
\end{equation}
can be expressed as
\begin{equation}
x^1=x^1_k(s,t), ~~~x^2=x^2_k(s,t),~~~x^3=x^3_k(s,t),
\end{equation}
which represent the world surfaces of $N$ moving isolated singular
strings $L_k$ with string parameter $s$ $(k=1,2,\cdots,N)$. These
singular strings solutions are just the phase singularities
solutions in three dimensions space.

In $\delta$-function theory\cite{deltfun}, one can obtain in three
dimensions space
\begin{equation}
\delta^2(\vec{\phi})=\sum_{k=1}^N \beta_k
\int_{L_k}\frac{\delta^3(\vec{x}-\vec{x}_k()s)}{|D(\frac{\phi}{u})|_{\Sigma_k}}ds,\label{delta3}
\end{equation}
where
$$D(\frac{\phi}{u})|_{\Sigma_k}=\frac{1}{2} \epsilon^{jk}\epsilon_{mn}\frac{\partial\phi^m}{\partial
u^j}\frac{\partial\phi^n}{\partial u^k},$$ and $\Sigma_k$ is the
$k$th planar element transverse to $L_k$ with local coordinates
$(u^1,u^2)$. The $\beta_k$ is the Hopf index of $\phi$ mapping,
which means that when $\vec{x}$ covers the neighborhood of the zero
point $\vec{x}_k(s)$ once, the vector field $\phi$ covers the
corresponding region in $\phi$ space $\beta_k$ times. Meanwhile the
direction vector of $L_k$ is given by
\begin{equation}
\frac{dx^i}{ds}|_{x_k}=\frac{D^i(\phi/x)}{D(\phi/u)}|_{x_k}.\label{direction}
\end{equation}
Then from Eq.(\ref{delta3}) and Eq.(\ref{direction}) one can obtain
the inner structure of $\omega^i$:
\begin{equation}
\omega^i=\sum_{k=1}^N W_k
\int_{L_k}\frac{dx^i}{ds}\delta^3(\vec{x}-\vec{x}_k(s))ds,\label{topo}
\end{equation}
where $W_k=\beta_k\eta_k$ is the winding number of $\vec{\phi}$
around $L_k$, with $\eta_k=sgn D(\phi/u)|_{\vec{x}_k}=\pm1$ being
the Brouwer degree of $\phi$ mapping. The sign of Brouwer degrees
are very important, for the case of vortices phase singularities,
the $\eta_k=+1$ corresponds to the vortex, and $\eta_k=-1$
corresponds to the antivortex. Hence the topological charge of the
phase singularity line $L_k$ is\cite{feild}
\begin{equation}
Q_k=\int_{\Sigma_k}\omega^id\sigma_i=W_k.
\end{equation}

The results in this section show us the topological inner structure
of the topological current density $\omega^i$. The topological
charge of the $k$th phase singularity line in three dimensions can
be expressed by the topological numbers: $Q_k=W_k=\beta_k\eta_k$.

 \section{Topological aspect on Knotted phase singularity lines}

An important case is that the phase singularity lines in three
dimensions space from closed and knotted curves. Topology has play a
very important role in understanding these knot configurations, so
it is necessary to study the topology in the knotted phase
singularity lines. In order to do that, we define a the helicity
integral\cite{heli}
\begin{equation}
H=\frac{1}{4\pi^2}\int \vec{v}\cdot\nabla\times\vec{v}d^3x,
\end{equation}
this is an important topological knot invariant and it measures the
linking of the phase singularity lines. From the
Eq.(\ref{vorticity}), the helicity integral can be changed as
\begin{equation}
H=\frac{1}{2\pi}\int \vec{v}\cdot\vec{\omega}d^3x.\label{helint}
\end{equation}
Substituting Eq. (\ref{topo}) into Eq. (\ref{helint}), one can
obtain
\begin{equation}
H=\frac{1}{2\pi}\sum_{k=1}^{N}W_k \int_{L_k}\vec{v}\cdot
d\vec{x}\label{helint1},
\end{equation}
when these phase singularities are closed and knotted lines, i.e., a
family of knots $\xi_k (k=1, 2, \ldots, N)$, Eq. (\ref{helint1})
becomes
\begin{equation}
H=\frac{1}{2\pi}\sum_{k=1}^{N}W_k \oint_{\xi_k}\vec{v}\cdot
d\vec{x}.
\end{equation}

It is well known that many important topological numbers are related
to a knot family such as the self-linking number and Gauss linking
number. In order to discuss these topological numbers of knotted
phase singularity lines, we define Gauss mapping:
\begin{equation}
\vec{m}: S^1 \times S^1 \rightarrow S^2,
\end{equation}
where $\vec{m}$ is a unit vector
\begin{equation}
\vec{m}(\vec{x},
\vec{y})=\frac{\vec{y}-\vec{x}}{|\vec{y}-\vec{x}|},\label{unit}
\end{equation}
where $\vec{x}$ and $\vec{y}$ are two points, respectively, on the
knots $\xi_k$ and $\xi_l$ (in particular, when $\vec{x}$ and
$\vec{y}$ are the same point on the same knot $\xi$ , $\vec{n}$ is
just the unit tangent vector $\vec{T}$ of $\xi$ at $\vec{x}$ ).
Therefore, when $\vec{x}$ and $\vec{y}$ , respectively, cover the
closed curves $\xi_k$ and $\xi_l$ once, $\vec{n}$ becomes the
section of sphere bundle $S^2$. So, on this $S^2$ we can define the
two-dimensional unit vector $\vec{e}=\vec{e}(\vec{x}, \vec{y})$.
$\vec{e}$, $\vec{m}$ are normal to each other, i.e. ,
\begin{eqnarray}
&&\vec{e}_1\cdot\vec{e}_2=\vec{e}_1\cdot\vec{m}=\vec{e}_2\cdot\vec{m}=0,
\nonumber\\&&\vec{e}_1\cdot\vec{e}_1=\vec{e}_2\cdot\vec{e}_2=\vec{m}\cdot\vec{m}=1.
\end{eqnarray}
In fact, the velocity field $\vec{v}$ can be decomposed in terms of
this two-dimensional unit vector $\vec{e}$:
$v_i=\epsilon_{ab}e^a\partial_i e^b-\partial_i\theta$, where
$\theta$ is a phase factor\cite{quanvortex}. Since one can see from
the expression $\vec{\omega}=\frac{1}{2\pi}\nabla\times \vec{v}$
that the $(\partial_i\theta)$ term does not contribute to the
integral $H$, $v_i$ can in fact be expressed as
\begin{equation}
v_i=\epsilon_{ab}e^a\partial_ie^b.
\end{equation}
Substituting it into Eq.(14), one can obtain
\begin{equation}
H=\frac{1}{2\pi}\sum_{k=1}^{N}W_k
\oint_{\xi_k}\epsilon_{ab}e^a(\vec{x},
\vec{y})\partial_ie^b(\vec{x}, \vec{y})dx^i.\label{hel}
\end{equation}
Noticing the symmetry between the points $\vec{x}$ and $\vec{y}$ in
Eq.(\ref{unit}), Eq.(\ref{hel}) should be reexpressed as
\begin{equation}
H=\frac{1}{2\pi}\sum_{k, l=1}^N W_k W_l \oint_{\xi_k}
\oint_{\xi_l}\epsilon_{ab}\partial_i e^a\partial_j e^bdx^i\wedge
dy^j.\label{hel2}
\end{equation}
In this expression there are three cases: (1) $\xi_k$ and $\xi_l$
are two different phase singularities $(\xi_k\neq\xi_l)$, and
$\vec{x}$ and $\vec{y}$ are therefore two different points
$(\vec{x}\neq\vec{y})$; (2) $\xi_k$ and $\xi_l$ are the same phase
singularities $(\xi_k=\xi_l)$, but $\vec{x}$ and $\vec{y}$ are two
different points $(\vec{x}\neq\vec{y})$; (3) $\xi_k$ and $\xi_l$ are
the same phase singularities $(\xi_k=\xi_l)$, and $\vec{x}$ and
$\vec{y}$ are the same points $(\vec{x}=\vec{y})$. Thus,
Eq.(\ref{hel2}) can be written as three terms:
\begin{eqnarray}
&&H=\sum_{k=1(k=l, \vec{x\neq}\vec{y})}^N \frac{1}{2\pi}W_k^2
\oint_{\xi_k} \oint_{\xi_k} \epsilon_{ab} \partial_i e^a\partial_j
e^b dx^i \wedge dy^j \nonumber\\&&+\frac{1}{2\pi}\sum_{k=1}^N W_k^2
\oint_{\xi_k} \epsilon_{ab} e^a\partial_i e^bdx^i
\nonumber\\&&+\sum_{k, l=1(k\neq l)}^N \frac{1}{2\pi}W_k W_l
\oint_{\xi_k} \oint_{\xi_l} \epsilon_{ab}
\partial_i e^a\partial_j e^b
dx^i \wedge dy^j.\label{hel3}
\end{eqnarray}
By making use of the relation
$\epsilon_{ab}\partial_ie^a\partial_je^b=\frac{1}{2}\vec{m}\cdot(\partial_i\vec{m}\times\partial_j\vec{m})$\cite{rela},
the Eq.(\ref{hel3}) is just
\begin{eqnarray}
H&=&\sum_{k=1(\vec{x}\neq\vec{y})}^N \frac{1}{4\pi}W_k^2
\oint_{\xi_k}\oint_{\xi_k} \vec{m}^*(dS) \nonumber \\ &&
+\frac{1}{2\pi}\sum_{k=1}^N W_k^2 \oint_{\xi_k} \epsilon_{ab}e^a
\partial_ie^bdx^i \nonumber \\ &&
+\sum_{k, l=1(k\neq l)}^N \frac{1}{4\pi} W_k W_l
\oint_{\xi_k}\oint_{\xi_l} \vec{m}^*(dS),\label{hel4}
\end{eqnarray}
where
$\vec{m}^*(dS)=\vec{m}\cdot(\partial_i\vec{m}\times\partial_j\vec{m})dx^i\wedge
dy^j(\vec{x}\neq\vec{y})$ denotes the pullback of the $S^2$ surface
element.

In the following we will investigate the three terms in the
Eq.(\ref{hel4}) in detail. Firstly, the first term of
Eq.(\ref{hel4}) is just related to the writhing number\cite{writnum}
$Wr(\xi_k)$ of $\xi_k$
\begin{equation}
Wr(\xi_k)=\frac{1}{4\pi}\oint_{\xi_k}\oint_{\xi_l}
\vec{m}^*(dS).\label{writhing}
\end{equation}
For the second term, one can prove that it is related to the
twisting number $Tw(\xi_k)$ of $\xi_k$
\begin{eqnarray}
\frac{1}{2\pi}\oint_{\xi_k}\epsilon_{ab}e^a\partial_ie^bdx^i
&&=\frac{1}{2\pi}\oint_{\xi_k}(\vec{T}\times\vec{V})\cdot
d\vec{V}\nonumber\\&&=Tw(\xi_k),\label{twisting}
\end{eqnarray}
where $\vec{T}$ is the unit tangent vector of knot $\xi_k$ at
$\vec{x}$ ($\vec{m}=\vec{T}$ when $\vec{x}=\vec{y}$) and $\vec{V}$
is defined as
$e^a=\epsilon^{ab}V^b(\vec{V}\perp\vec{T},\vec{e}=\vec{T}\times\vec{V})$.
In terms of the White formula\cite{writfor}
\begin{equation}
SL(\xi_k)=Wr(\xi_k)+Tw(\xi_k),\label{self}
\end{equation}
we see that the first and the second terms of Eq.(\ref{hel4}) just
compose the self-linking numbers of knots.

Secondly, for the third term, one can prove that
\begin{eqnarray}
&&\frac{1}{4\pi}\oint_{\xi_k}\oint_{\xi_l}
\vec{m}^*(dS)\nonumber\\&&=\frac{1}{4\pi}\epsilon^{ijk}\oint_{\xi_k}dx^i\oint_{\xi_l}dy^j
\frac{(x^k-y^k)}{\|\vec{x}-\vec{y}\|^3}\nonumber\\&&=Lk(\xi_k,\xi_l)~~(k\neq
l),\label{linking}
\end{eqnarray}
where $Lk(\xi_k,\xi_l)$ is the Gauss linking number between $\xi_k$
and $\xi_l$\cite{writnum}. Therefore, from Eqs.(\ref{writhing}),
(\ref{twisting}), (\ref{self}) and (\ref{linking}), we obtain the
important result:
\begin{equation}
H=\sum_{k=1}^N W_k^2 SL(\xi_k)+\sum_{k, l=1(k\neq
l)}^NW_kW_lLk(\xi_k, \xi_l).
\end{equation}
This precise expression just reveals the relationship between $H$
and the self-linking and the linking numbers of the phase
singularity knots family\cite{writnum}. Since the self-linking and
the linking numbers are both the invariant characteristic numbers of
the phase singularity knots family in topology, $H$ is an important
topological invariant required to describe the linked phase
singularities in wave fields.

 \section{Conclusion}
In the present study, by making use of the $\phi$-mapping
topological current theory, the topology of the phase singularity in
wave fields is studied. Firstly, we obtain the inner structure of
the phase singularities in two and three dimensional space. The
phase singularity objects have been found at the every zero point of
the wave function $\psi$ under the condition that the Jacobian
determinate $D^i(\phi/x)\neq0$. One shows that the topological
charge (strength of the phase singularity) of the phase singularity
objects are determined by the topological numbers: Hopf indices and
Brouwer degrees. Secondly, we have studied the topological invariant
of the knotted phase singularity lines in three dimensional space in
details. This topological invariant can be expressed as the sum of
all the self-linking and all the linking numbers of the knotted
phase singularity lines.

Finally, it should be pointed out that in the present paper, when we
discussed the topological properties of the phase singularities, the
condition $D^i(\phi/x)\neq0$ must be satisfied. Now the question is
coming, when this condition fails, what will happen about the phase
singularity objects? The answer is related to the evolution of the
phase singularity objects\cite{evolution}. The topological object
will generate, annihilate, split, or merge when the condition fails
, and these dynamics properties of the phase singularity objects
will be discussed in further work.

\begin{acknowledgments}
This work was supported by the National Natural Science Foundation
of China.
\end{acknowledgments}

 \end{document}